\documentclass[10pt]{article}  
\usepackage{graphicx}  
\usepackage{amssymb}
\usepackage{amsmath}   
\usepackage{slashed}
\begin{document}
	
\title{Holographic Corrections to the Veneziano Amplitude}
\date{\today}

\author{Adi Armoni$^{\,1}$, Edwin Ireson$^{\,1}$
\\
\textit{$^{\,1}$ Department of Physics, Swansea University,}\\ \textit{Swansea SA28PP, United Kingdom}}

\maketitle
\begin{abstract}
We propose a holographic computation of the $2\rightarrow 2$ meson scattering in a curved string background, dual to a QCD-like theory. We recover the Veneziano amplitude and compute a perturbative correction due to the background curvature. The result implies a small deviation from a linear trajectory, which is a requirement of the UV regime of QCD.
\end{abstract}

\maketitle
\section{Introduction}

 The Veneziano amplitude, aimed at describing meson scattering, marks the birth of string theory \cite{Veneziano:1968yb}. While the Veneziano amplitude exhibits several attractive properties, such as the duality between the s-channel and the t-channel, and a phenomenological success in the Regge regime, it also suffers from an undesired exponential behaviour in the high energy regime.

The properties of the Veneziano amplitude are closely related to the phenomenon of confinement. Makeenko and Olesen \cite{Makeenko:2009rf} showed that the amplitude can be reproduced from large-$N$ QCD, by representing the amplitude as a sum over Wilson loops. A crucial unrealistic assumption, is that all Wilson loops (even small loops) admit an area law. 

In a recent attempt to derive the Veneziano amplitude from large-$N$ QCD \cite{Armoni:2015nja}, the sum over Wilson loops of \cite{Makeenko:2009rf} was written by using holography \cite{Maldacena:1997re} as a sum over string worldsheets. 
The sum includes all Wilson loops that pass via the positions of the mesons.
The Veneziano amplitude was obtained under the assumption that Wilson loops are saturated by flat space configurations that sit on the IR cut-off (hence leading to an area law). The flat space approximation can be achieved by bringing the IR cut-off close to the UV cut-off (the space boundary).

Thus one can attribute the failure of the Veneziano amplitude in the high energy regime to the flat space approximation, which is identical to the assumption that all Wilson loops admit an area law. It is therefore desired to accommodate small Wilson loops. In the dual string theory it means that one has to calculate string amplitudes in curved space. Even if that could be done, an exact dual of large-$N$ QCD is not known, hence it is not clear which non-linear sigma model should be used.

In this paper we propose a method to improve the scattering amplitude, by incorporating curvature effects from the dual geometry. In particular we study the effect of worldsheet fluctuations in the vicinity of the IR cut-off.  As the worldsheet fluctuates it probes part of the UV geometry. To this end we include in the string sigma model an interaction term between the flat 4d coordinates and extra dimensions and calculate a two-loop perturbative correction to the propagator. While we carry out the calculation by using Witten's model of the dual of Yang-Mills theory \cite{Witten:1998zw}, we believe that the sigma model we use characterizes generic confining holographic models. The result of the calculation is a correction to the Veneziano amplitude in the form of a deviation from a linear Regge trajectory $\alpha(s)\sim s^{(1-\rho^3 \log^2 s)}$, with $ \rho^3 \log^2 s \ll 1$.

 Our approach is distinct from earlier studies of Pomeron scattering using the AdS/CFT correspondence \cite{Polchinski:2001tt}. The novelty, apart from considering meson scattering (open string amplitude), is that we include a perturbative correction to the $X^\mu$ propagator. 

Our conclusion about the non-linearity of the Regge trajectory at small angular momenta is similar to that of previous studies \cite{Imoto:2010ef,Sonnenschein:2016pim}. It is consistent with empirical data \cite{Sonnenschein:2016pim}.

\section{Setting up the duality} 
The starting point of this framework is the aforementioned identity involving a sum over all sizes of Wilson loops. To compute this four-point function we make use of the Worldline formalism, namely the following equality for the fermion determinant of $SU(N_c)$ QCD with $N_f$ flavours, in terms of expectation values of super-Wilson loops of length $T$:

\begin{align}
\mathcal{Z}=\int DA \exp(-S_{\rm YM}) &\left( {\det} \left( i \slashed{D} \right) \right) ^{N_f}, \,\,\,\,
\left( {\det} \left(i \slashed{D} \right)\right)  ^{N_f} = \exp\left(-\frac{N_f}{2}\text{Tr}\int_0^\infty \frac{dT}{T}  \left<\mathcal{W}_T[A]\right>\right)\text{ , }
\\ &\left<\mathcal{W}_T[A]\right>=\int DxD\psi e^{-\frac{1}{2}\int_0^T d\tau (\dot{x}^2 + \psi \cdot \dot{\psi})}e^{i\int_0^T d\tau (\dot{x}\cdot A - \frac{1}{2}\psi\cdot F \cdot \psi )}.\nonumber
\end{align}

This exponential can then be expanded in powers of $N_f/N_c$, so that in the 't Hooft large $N_c$ limit only the linear term in Wilson loops needs to be considered. We will use this approximation (the so-called ``quenched approximation'') for the following computation of a generic meson 4-point function \cite{Armoni:2015nja}:
\begin{align}
\left<\prod\limits_{i=1}^4 q\bar{q}(x_i)\right>={1\over \cal Z} \prod\limits_{i=1}^4 {\delta \over \delta J(x_i)} \int DA\exp(-S_{\rm YM}) \times  
\left(-\frac{N_f}{2}\text{Tr}\int_0^\infty \frac{dT}{T}  \left<\mathcal{W}_T[A,J]\right> \right) | _{J=0}
\end{align}
with $\left<\mathcal{W}_T[A,J]\right>$ the worldline action of a theory coupled to a meson source $S_J = \int d^4 x \, J\bar q q$ \cite{Armoni:2011qv}. Taking functional derivatives in terms of $J$ creates delta functions constraining the various loops at hand to pass through the points where the operators are inserted. We therefore arrive at the following formal expression for the scattering amplitude 
\begin{align}
\left<\prod\limits_{i=1}^4 q\bar{q}(x_i)\right>={1\over \cal Z} \int DA\exp(-S_{\rm YM}) \times  
 \left(-\frac{N_f}{2}\text{Tr}\int_0^\infty \frac{dT}{T}  \left<\mathcal{W}_T[A]\right>|_{x_1,x_2,x_3,x_4}\right),
\end{align}
where the sum over all Wilson loops has been imposed to include only those loops passing through the 4 points where the meson operators have been inserted.

The gauge/gravity duality implies that, for the field theory on the boundary, a generic Wilson loop's expectation value is related to the expectation value of a string worldsheet hanging from the loop on the boundary down into the bulk. More precisely, a single Wilson loop expectation value is obtained as the saddle of a sum over string worldsheets, as explained in section 3 of \cite{Drukker:1999zq}. It was then proposed that the contribution of ``quenched flavours'' to the gauge theory partition function is dual to a sum over all string worldsheet with a topology of a disk that terminate on the boundary of the AdS space \cite{Armoni:2008jy}
\begin{align}
  \int DA\exp(-S_{\rm YM}) \times \left(-\frac{1}{2}\text{Tr}\int_0^\infty \frac{dT}{T}  \left<\mathcal{W}_T[A]\right> \right)  = \int [DX] [Dg] \exp\left( -{1\over 2\pi \alpha '}{\int d^2\sigma\, G_{MN} \partial_\alpha X^M \partial _\beta X^N g^{\alpha \beta}} \right ) \,.
\end{align}
This identification is possible because we have fixed a flat worldsheet gauge $ g^{\alpha \beta}= \delta ^{\alpha \beta}$, we are not interested in higher genus contributions. Then, schematically, the above equality holds because both path integration measures on either side sum up area-behaved integrands over all possible shapes and sizes of their boundaries. We are therefore able to write the following expression for the amplitude\cite{Armoni:2008jy}

\begin{align}
\mathcal{A}\left(k_{i=1\dots4} \right) =\oint \prod\limits_{i=1}^4d\sigma_i \int [DX]W e^{ik_i^\mu X_\mu(\sigma_i)} \times
\exp\left( -{1\over 2\pi \alpha '}{\int d^2\sigma\, G_{MN} \partial_\alpha X^M \partial ^\alpha X^N}\right) 
,\label{stringamp}
\end{align}
where $W e^{ik_i^\mu X_\mu(\sigma_i)}$ are meson vertex operators in the dual string theory, inserted at different points on the worldsheet, each associated to a momentum $k_i$. $W$ encodes details of spin ($W=1$ for the tachyon and $W=\epsilon \partial _\sigma  X$ for a vector for the bosonic string), which only really affects the intercept $\alpha_0$ of the Regge trajectory, $\alpha(s) = \alpha_0 +\alpha ' s$. As we shall see the calculated  perturbative correction is not valid near $s=0$. In addition, there may be additional subleading phenomena contributing to a shift of the intercept, hence we can make no statements thereupon. The expectation values of the vertex operators are taken with respect to the Polyakov-type non-linear sigma model action over the string worldsheet into a curved space. Choosing the correct background is crucial in this matter. Since we have done away with flavour degrees of freedom via the worldline formalism, we only need to pick a space dual to pure Yang-Mills. For our purposes, we take Witten's background of back-reacted $D4$-branes compactified on a thermal circle \cite{Witten:1998zw}, whose dual, while not pure Yang-Mills, possesses enough similarities (confinement and a mass gap) that we can hope to make generic arguments thereupon. The chief property of this space is its metric $G$, inducing the following space-time line element where $X^\mu$ $(\mu=0,1,2,3)$ are boundary space-time coordinates, $\tau$ the compact direction, $U$ the AdS direction and the remaining 4 coordinates parametrise a sphere:

\begin{align}
 ds^2&=g(U)\left(d X^2+\text{d$\tau $}^2 f(U)\right)+\frac{1}{g(U)}\left( \frac{d U^2 }{f(U)}+ U^2d\Omega ^2 _4\right) \nonumber\\
 &\text{where }g(U)=\left(\frac{U}{R}\right)^{3/2}\textbf{, }f(U)=1-\frac{U_{\text{KK}}^3}{U^3}
\label{action}.
 \end{align}

It admits a metric singularity at the point $U=U_{KK}$, where the space has a horizon. As is usual, string worldsheets hanging from large loops (of size comparable to the horizon position) will accumulate on the horizon, providing an area-law scaling of their expectation value, rather than a perimeter-law, this is the usual tell-tale sign that confinement has taken place. Then, provided the strings do not have to go very far from the boundary to the horizon, the classical saddle of this action is a string whose geometry is mostly flat, spread out on the horizon itself. By pushing $U_{KK}$ close to infinity, this condition is broadly satisfied, turning the classical saddle of the path integral into a mostly-flat worldsheet. This was the claim proposed in a previous work \cite{Armoni:2015nja}, from which the 4-point function we wish to compute fairly naturally reproduces the Veneziano amplitude as the behaviour associated to flat open-string scattering. 

These final steps relied on many broad assumptions, namely that we ignore effects coming from the compact directions (justified through the hierarchy of scales at hand), from the fermionic degrees of freedom (justified by spacetime supersymmetry breaking, and by the insertion of purely bosonic operators), to impose that almost all worldsheets in the sum are heavily accumulating on the horizon, discarding those that do not and ignoring the contribution of the edges of those that do (justified by the near-infinite size of $U_{KK}$), and to assume no string quantum corrections (both genus/ghost and $\alpha^\prime$). Our goal is to relax the latter, to allow string tension corrections to the computation at hand, which in physical terms corresponds to letting the string fluctuate around its classical position and experience the $U$ direction curvature and (it will be shown) the compact direction $\tau$.

\textit{Setting up an expansion.}--- From the Polyakov action using the metric shown in Eq.(\ref{action}), we wish to create a perturbation series for values of the AdS coordinate $U$ close to the horizon position, $U_{KK}$, which for previously explained reasons should be thought of as a large length scale. The first step is to provide an adequate parametrisation of the space, in which the coordinate singularity at the horizon is removed. Typically a set of Kruskal-like coordinates are desirable, as has been done in similar computations in the past \cite{Greensite:1999jw}, but unlike that particular example, they are non-trivial to compute in the case at hand, it is unclear whether a good new radial coordinate can be analytically computed over the entire AdS region. We will therefore assume that the string fluctuations are of small amplitude and only experience a small range of the target space curvature: in practice this means we can perform an expansion of the offending metric elements for values of $U$ parametrically close to $U_{KK}$ and compute near-horizon Kruskal coordinates, rather than try to globally define them.

At this cost, the operation can be done and results in the following coordinate definitions: introducing $\lambda=\frac{U_{KK}}{R}$ as an arbitrary constant, we define new coordinates by
\begin{align}
\frac{U}{U_{KK}}=1+\frac{u^2}{U^2_{KK}}, \,\,\, \Upsilon = ue^{i\sqrt{\frac{9 \lambda ^3}{4 U_{KK}^2}}\tau}e^{\frac{u^2}{4U_{KK}^2}} \, ,
\end{align}
all power series in $u/U_{KK}$ can now be rewritten as series in $|\Upsilon / U_{KK}|$, as the relation above is easily invertible. This produces a string action in a very appropriate form for interaction expansions:
\begin{align}
{\cal L}=\lambda^{3/2}\partial_\alpha X^\mu \partial^\alpha X_\mu + \frac{4}{3\lambda^{3/2}}\partial_\alpha\Upsilon^\dagger\partial^\alpha\Upsilon 
 + \frac{6}{T U^2_{KK}}\Upsilon^\dagger \Upsilon+\frac{3\lambda^{3/2}}{2U_{KK}^2}\Upsilon^\dagger \Upsilon \partial_\alpha X^\mu \partial^\alpha X_\mu + \dots
\end{align}

The last term is the direct result of this process, an interaction term between $\Upsilon$ and $X^\mu$, but we have also gained a mass term for this AdS coordinate: it has come from a Jacobian factor multiplying the invariant path integration measure. Note that it is parametrically small as it depends both on the string tension $T$ and $U_{KK}$, also that this coordinate redefinition characterises not only vertical motion of the worldsheet but also motion in the $\tau$ direction. We keep ignoring effects from the additional compact coordinates, seen as an unfortunate artefact of the string background. We also ignore the effects of supersymmetry: certainly, the string theory has explicit broken space-time supersymmetry, but one does not expect to see it explicitly from the worldsheet, and so we should include fermions in the action. However, we note that, firstly, we are inserting purely bosonic operators in the path integral, so we need to focus on interactions featuring those bosonic fields, secondly, that such terms connecting bosonic and fermionic fields in a generic non-linear sigma model ($\Gamma_{\mu \nu \rho}\bar{\psi}^\mu \slashed{\partial}X^\nu \psi^\rho)$ vanish in our case, through specific properties of the Christoffel symbol in this background. For this reason the coupling to RR background fields is also expected to be subleading as they couple primarily to the worldsheet fermions.

The issue of the dilaton coupling to the worldsheet Ricci tensor (Fradkin-Tseytlin coupling) is more subtle. In our setup we consider the kinematical regime (Regge regime) where the sum over string worldsheets is dominated by large area and flat-space worldsheets, calculated in the vicinity of $U=U_{KK}$. For such worldsheets we may take the approximation  $\int d^2 \sigma \, \Phi R \approx \int d^2 \sigma \, \Phi (U_{KK}) R$, where the contribution is similar to the flat space case and the ``area term'' $\int \partial X \partial X$ is expected to be the dominant contribution to the action.

We can now proceed to perform perturbation theory around the $U=U_{KK}$ "vacuum" of the effective field theory described by this Lagrangian.

\section{Evaluation of loop integrals}

This action now having a well-defined vacuum to expand around, we set out to compute the quantity we defined previously in Eq.(\ref{stringamp}). From general considerations on the tower-like structure of string excited states we argue that considering vertex operators of the form $e^{ i k_i \cdot X(\sigma_i)}$ should be informative enough (again noting that we are not interested in the intercept). The purely classical approximation immediately reproduces the Beta function behaviour of the amplitude.By adding these operators in the action as a current $\mathcal{J}(\sigma)= \sum\limits_{i=1}^4 k_i \cdot X(\sigma) \delta(\sigma-\sigma_i)$, this essentially maps the computation onto the calculation of the partition function around a non-zero current (suitably normalised by that without any sourcing). As a result we are computing "vacuum" diagrams with no in/out states, but not vacuum bubbles-- rather, all those whose legs are stopped at the positions where the current is being introduced. These effectively 1-leg vertices are dimensionful, thus the more legs a diagram has, the higher its order is in our series expansion, the dimensions being soaked up by powers of $1/U_{KK}$. Thus we will focus on worldsheet 2-point corrections.

Now, given the form of the interaction at hand, and the fact we are dealing with a 2D QFT, we will frequently have to deal with logarithmically divergent integrals, starting with the propagators, and thus have to work to remove regulator poles but also logarithms of vanishing or asymptotic quantities. For this purpose we will use analytic regularisation of the propagators, along with a convenient version of the $\overline{MS}$ scheme. Once this is done, our framework has the particularity of making a large class of diagrams, that we can write with this new interaction vertex, trivial in a sense. The lowest order diagrams in our expansion, the 2-point 1-loop graphs with no momentum transfer, factorise into a number of "vacuum bubble"-like integrals times an overall propagator, our regulation scheme makes such diagrams finite. This corresponds to, equivalently, a finite shift of the wavefunction normalisation, or a finite shift of the string tension $T$. As the bare value of $T$ is tunable, we should consider that at the end of the procedure, once all the finite shifts in tension at all orders have been applied, the new, effective string tension is the physical QCD one. We will hereafter always refer to the effective string tension when writing $T$ or $\alpha^\prime$.

Next, we are brought to study a two-loop correction to the two-point amplitude (curiously lower order in $U_{KK}$ than lower-loop, higher-point amplitudes by dimensional analysis) taking the form of a "sunset" diagram,
\begin{figure}[h]
\centering
\includegraphics[width=0.7\textwidth]{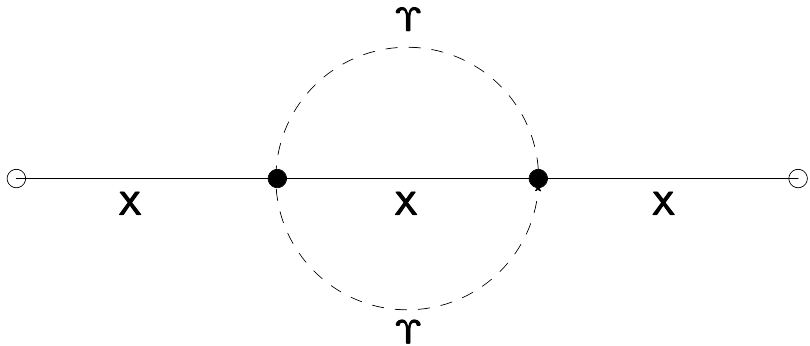}
\caption{The first diagram leading to qualitative corrections to the Veneziano amplitude.}
\label{2loop}
\end{figure}
corresponding to the following integral

\begin{align}
I= \sum\limits_{i<j}k_i\cdot k_j\frac{81(\pi\alpha^\prime)^3\lambda^{3/2}}{64 U_{KK}^4}\times
{\Large \int} \frac{d^2p d^2 q_1 d^2 q_2}{(2\pi)^6} 
\frac{e^{i(p\cdot \left(\sigma_i - \sigma_j \right) )}(p\cdot (q_1+q_2))^2}{p^4(p-q_1-q_2)^2(q_1^2+m^2)(q_2^2+m^2)}
\end{align}

This integral is not trivial-- especially given that only two out of the three internal lines are massive. Some work has been done on the analytics of similar diagrams \cite{Smirnov:2012gma}, but not using a regulation scheme adapted to logarithmic divergences as is our case. However, at this point we use the fact that the mass in question is parametrically small: it is of the same order as the coupling constant of our interaction term. Since the massless diagram is well-defined, we should therefore think of it as truly being the lead source of qualitative corrections to the Veneziano regime, the mass inducing higher-order corrections. At this cost, the diagram is then well-suited to treatment in our set-up, and we obtain a finite value for it:

\begin{equation}
I= -\frac{1}{8\pi^3} \frac{27\lambda^{3/2}}{64 T^3 U_{KK}^4}\sum_{i<j} k_i\cdot k_j \left(\log^3\left(\left(\sigma_i-\sigma_j\right)^2\right) \right) .
\end{equation}

The computation closes as we insert this result in the overall expectation value and averaging over all possible insertion positions described above in Eq.(\ref{stringamp}). We encapsulate the expansion parameter and the numerical factors by defining, $\rho^3=\frac{1}{4\pi^2}\frac{27\lambda^{3/2}}{64 T^3 U_{KK}^4}$, and we will write scalar products of space-time momenta $k_i\cdot k_j$ as Mandelstam variables $s,t$: $s=-k_1\cdot k_2\,\, ,\,\, t=-k_1\cdot k_3$. Systematically these variables come with a factor of $\alpha^\prime$, which we will absorb in the definition of these variables to make them dimensionless.

Then the amplitude becomes
\begin{align}
\mathcal{A}(s,t)=\int_{0}^{1} dz z^{-s-1} (1-z)^{-t-1}\times
\left(1+\rho^3 \left(s \log^3 (z) + t \log^3(1-z) \right) \right),
\end{align}
which, one can quickly recognize, has a leading term corresponding precisely to the Veneziano Beta-like functional form, with extra logarithmic corrections. The corrective integrals can be computed explicitly and give third derivatives of the Beta function,

\begin{equation}
\mathcal{A}(s,t)=\left( 1-\rho^3\left( s\frac{\partial^3}{\partial s^3}+ t\frac{\partial^3}{\partial t^3}\right) \right) B(-s,-t).
\end{equation}
This expression is actually composed of many terms of varying relevance due to properties of Beta and associated functions.

\section{Consequences on the Regge trajectory}
To recover the incidence of these corrections on the Regge behaviour, we will express the amplitude in an approximate form, in the so-called Regge regime.

\begin{figure}[h!]
\centering
\includegraphics[width=0.7\textwidth]{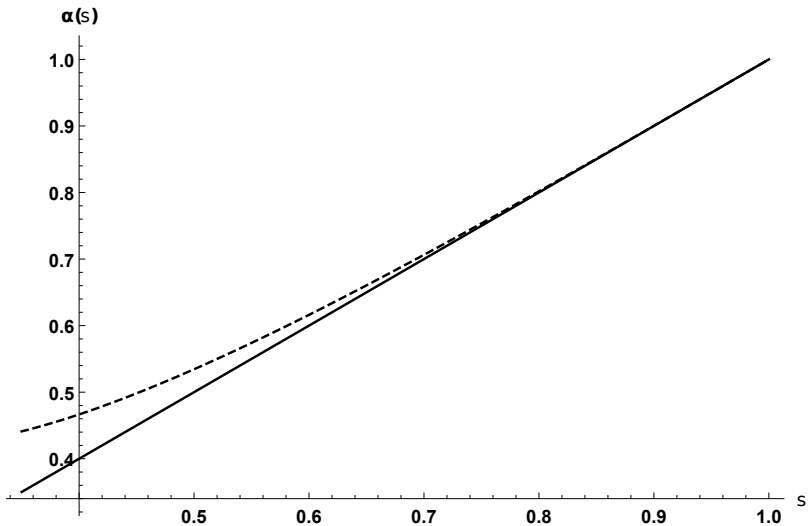}
\caption{Comparing the new Regge trajectory (dashed) to linear behaviour (solid), for $\rho^3=0.2$}
\label{regge2}
\end{figure}

 We then take the limit $|s|\gg |t|$,  $|t|$ fixed. We expect to have $B(-s,-t)\propto \left(-s \right) ^{t}$ (ignoring the purely $t$-dependent part of the amplitude, which contains the pole at zero in $t$). This form is still valid in the elastic collision regime, as long as one moves a little away from the positive real $s$-axis, as is routinely done in such matters. This ensures that the influence of the poles and the zeroes of the function do not disrupt this approximate form too much. The nature of the Beta function means that this limit is quite lax, $|s|$ need not be large in magnitude for it to be adequate, simply comparatively larger than $|t|$. Remarking that every $s$ derivative lowers the order of the asymptotic behaviour of the function, the main contributions come from the $t$ derivatives. Generically either of these derivatives involve the Polygamma functions $\{\Psi_{i\in \mathbb{N}}\}$, all of which vanish as powers of their argument asymptotically, save for the first one, $\Psi_0$, which diverges logarithmically. These terms will dominate the functional form of the correction in the limit we have chosen. With this prescription we find an approximate form for the amplitude 
\begin{equation}
\mathcal{A}(s,t)= \exp \left ( t(\log (-s) - \rho ^3 \log ^3 (-s) )\right ) \label{main-result}
\end{equation}
 and correspondingly a Regge trajectory of the form
\begin{equation}
\alpha(-s)\sim (-s)^{1-\rho^3 \log^2(-s)}
\end{equation}
where $\alpha(s)=s$ would be the leading order, purely classical string result. This is a satisfactory result, in that the Regge behaviour for values of $|s|<1$ starts to deviate away from the linear behaviour, in qualitatively similar ways as in previous efforts, namely, when plotted on the $(s,\alpha(s))$ plane, the curve bends towards the $\alpha(s)$ axis, as shown in Figure \ref{regge2}.

 It is divergent at the origin, but at this point our approximations no longer hold, as is the case generically in similar studies. This does not result in any predictions for the spectrum of the theory.

\section{Conclusions and outlook}

We have considered a computation of $2\rightarrow 2$ meson scattering using holography, recovering the Veneziano amplitude and computing a correction due to interactions between the spacetime coordinates and bulk coordinates $U$ and $\tau$, whose form we expect to be generic, this can be argued from properties of confining spacetimes.

Our main result is (\ref{main-result}), valid in the Regge regime where the tree level Veneziano amplitude is a good approximation. The correction we found affects the low energy regime (low $s$) of the Regge trajectory, as depicted in Fig.\ref{regge2}. Such a correction is required in QCD, since perfectly linear Regge trajectories mean a linear confining potential even at short distances. This is in conflict with asymptotic freedom.

It is encouraging to note that our results compare well with other studies and empirical data. For example, in ref.\cite{Sonnenschein:2016pim} the author finds a similar qualitative correction to the Regge trajectory in the low $s$ regime and claims that the empirical data of heavy meson spectra is better fitted by such a non-linear trajectory. Moreover, in ref.\cite{Imoto:2010ef} the authors calculate a correction to the meson spectrum due to the curvature of the background. They also find a qualitative non-linear behaviour similar to Fig.\ref{regge2} and fit it to the $\rho$-meson spectrum. Both cases plot $M^2$ as a function of $J$, the mirror of our graph, and get a bending towards $J$.
 
Encouraged by our results, we hope to apply our approach to other models and consider other processes such as meson-baryon scattering.

\textit{Acknowledgement.}--- We thank Timothy Hollowood, Zohar Komargodski and Carlos N\'{u}\~{n}ez for useful discussions and comments. 

\newpage

\providecommand{\href}[2]{#2}\begingroup\raggedright\endgroup

\end{document}